\documentstyle[preprint,prb,aps]{revtex}

\begin{document}
\draft

\title{Conformal Invariance in Periodic Quantum Chains}

\author{Rudolf A.\ R\"{o}mer and Bill Sutherland}

\address{Physics Department, University of Utah, Salt Lake City, UT 84112}

%\date{Feb.\ 2.\ 1993}
\date{\today}
\maketitle

\begin{abstract}
We show how conformal invariance predicts the functional form
of two-point correlators in one-dimensional periodic quantum
systems.
Numerical evidence for this functional form in a wide class of models
--- including long-ranged ones --- is given and it is shown how this
may be used to significantly speed up calculations of critical exponents.
\end{abstract}


%\narrowtext
%\tighten

%%%%%%%%%%%%%%%%%%%%%%%%%%%%%%%%%%%%%%%%%%%%%%%%%%%%%%%%%%%%%%%%%%%%%%%%
%
% Introduction
%
%%%%%%%%%%%%%%%%%%%%%%%%%%%%%%%%%%%%%%%%%%%%%%%%%%%%%%%%%%%%%%%%%%%%%%%%

Conformal invariance has been shown to be remarkably powerful in
predicting the universal long-distance behavior of low-dimensional
field theories and statistical systems both at the classical and the
quantum level \cite{p70,c87,isz88}.
Independent verification of conformal predictions by
finite-size studies was made possible by Cardy's \cite{c84}
use of conformal mappings to relate a problem in one geometry to one
in a finite-sized geometry. The numerical success of this approach
offers a convincing validation of the principle of conformal
invariance at critical points \cite{ds82,nb83}.

The main ideas leading to the conformal finite-size approach in
two-dimensions may be summarized as follows.
Under a conformal mapping $z\mapsto w(z)$, the two-point correlation
function of a scalar scaling operator $\phi(z)$ at a critical point
will transform covariantly, i.e.\
\begin{displaymath}
\FL
\langle \phi(z_{1},\bar{z}_{1}),\phi(z_{2},\bar{z}_{2}) \rangle=
|w'(z_{1})|^{x} |w'(z_{2})|^{x}
\langle \phi(w_{1},\bar{w}_{1}),\phi(w_{2},\bar{w}_{2}) \rangle,
\end{displaymath}
where $x$ is the scaling dimension of $\phi$.
A mapping of the entire $z$-plane onto the surface of a cylinder
of circumference $L$ is achieved by choosing the particular conformal
mapping \cite{c84}
\begin{displaymath}
w = \frac{L}{2\pi} \ln(z)
\end{displaymath}
We can then relate the two-point function of the scalar operator
$\phi(z)$ in the infinite plane, to the two-point function
of the operator $\phi(w)$, now evaluated in the cylinder geometry.
Putting $w=u + i v$, so that $u$ measures distance along the infinite
cylinder and $v$ is the periodic coordinate across, we may explicitly
calculate the two-point function to be \cite{c87}
\begin{equation}
\FL
\langle \phi(u_{1},v_{1}),\phi(u_{2},v_{2}) \rangle =
\frac{ (2 \pi / L)^{2x} }{
\left[
2 \cosh \frac{2\pi}{L}(u_{1}-u_{2}) - 2 \cos \frac{2\pi}{L}(v_{1}-v_{2})
\right]^{x}}
\label{eqn-cf}
\end{equation}
For large separation along the cylinder axis, i.e.\ assuming
$u_{1}-u_{2} \gg L$, the correlator exhibits an exponential
decay
\begin{displaymath}
\FL
\langle \phi(u_{1},v_{1}),\phi(u_{2},v_{2}) \rangle =
\left(\frac{2 \pi}{L}\right)^{2x}
e^{-(2\pi x/L) (u_{1}-u_{2})}
\end{displaymath}
and we may deduce the correlation length
\begin{equation}
\xi = \frac{L}{2\pi x}
\label{eqn-clsd}
\end{equation}
%The inverse correlation length $\xi^{-1}$ in turn is related
%to the ratio of the eigenvalues of the transfer matrix --- a finite
%matrix for the cylinder of finite circumference.

For one-dimensional quantum systems, we now reinterpret the coordinate
$u$ as running along the infinite chain of sites and the width $L$ in the $v$
direction as corresponding to the inverse temperature $\beta$, i.e.\ the
above reviewed 2D results may be applied also to an infinitely long
quantum chain at finite temperature.
This then opens the possibility of checking the predictions for the
critical exponents as obtained by (\ref{eqn-clsd}), by directly
studying the long-distance behavior of the correlation functions.

Analytic results for correlation functions, however, can only be
given for a rather select group of models
\cite{su7x,vt79,jmms80} and thus it is
at this point that numerical studies and simulations become
indispensable tools for the verification of conformal results.
Various methods for the numerically challenging problem of
calculating correlation functions have been devised over the
years \cite{b92}, but due to the complexity of the problem,
computations are still limited to lattice sizes of about $N<100$.
The extrapolation of the critical exponents from these finite-size
data usually involves taking the thermodynamic limit by
holding the density $d$ fixed as $N\rightarrow\infty$, as a first
step.
However, the correlation functions still exhibit an oscillatory
behavior \cite{vt79,jmms80}.
For the one-particle reduced density matrix $\rho(r)$, this behavior
leads to model-specific singularities in the momentum distribution $n(k)$.
Thus, the next step in extracting the long-distance behavior
should be to let the average inter-site (or inter-particle)
spacing $1/d$ go to zero.
The oscillations will then be pushed towards the origin.
(For small lattice sizes or calculations at fixed density
\cite{gdma86,kkb88,lc91}, where this limit is not practical,
one can instead consider a suitable averaging procedure,
leading to equivalent results.)

There is, however, another interpretation of the two-point function
(\ref{eqn-cf}). Most of the above numerical studies are performed not on
a long chain with free boundary conditions, but on a ring of length
N with periodic boundary conditions. Thus it seems natural, to
use the periodic coordinate $v$ to label the sites and
the coordinate $u$ to measure the time. In this picture, we just
`roll' the cylinder in a different way, i.e.\ we have a ring of length
$L=N$ and infinite extent in time.
The equal-time correlator may then be constructed from (\ref{eqn-cf})
by choosing $\Delta u=u_{1}-u_{2}=0$ and defining
$\Delta v=v_{1}-v_{2} \equiv r$.
The two-point function of (\ref{eqn-cf}) now reads as
\begin{equation}
\FL
\langle \phi(r),\phi(0) \rangle =
 \left(
  \frac{ 2 \pi }{ L }
 \right)^{2x}
 \left(
  \frac{ 1 }{ \sin \frac{\pi r}{L} }
 \right)^{2x},
\label{eqn-etcf}
\end{equation}
where we have suppressed the $u_{1},u_{2}$ arguments.
Thus we arrive at a prediction, not only of the critical exponents
of a given theory, but also of the analytic behavior of the
correlation functions for a finite-sized system on a ring.

Before we present numerical evidence for the above functional
form, let us examine in what limit this behavior is expected to
occur.
Because of the non-universal oscillations, we have true agreement
only for $1/d\rightarrow 0$.
This limit can easily be achieved by plotting the two-point function
as a function of $r/L$ on a fixed scale from $-1/2$ to $+1/2$.
We then expect the universal law (\ref{eqn-etcf}) to dominate as
$L\rightarrow\infty$.
(Note that for fermionic systems, we use the envelope of the two-point
function.)

The consequences of (\ref{eqn-etcf}) are impressive. First of all, we
have a much stronger test of conformal invariance.
After all, not only must the critical exponents agree, but also the complete
analytic behavior.
Let us now test our scaling law for the reduced one-particle density
matrix $\rho(r)$ and the spin-correlation $\langle S_{0} S_{r} \rangle$
in a variety of models.
For the $s=1/2$ Heisenberg antiferromagnet, it has been shown via bosonization
techniques \cite{lp75} that the singular behavior of
$\langle S_{0} S_{r} \rangle$, i.e.\
\begin{equation}
\langle S_{0} S_{r} \rangle \propto (-1)^{r} r^{-\eta}
\label{eqn-lp}
\end{equation}
is described by $\eta=1$. Recently, long standing conjectures
of possible logarithmic corrections have been verified by including
marginal operators in the above reviewed conformal finite-size analysis
\cite{n92}.
In Fig.~\ref{fig-af} we show a comparison of recent Monte-Carlo simulations
\cite{gdma86,kkb88,lc91} and the scaling law (\ref{eqn-etcf}). Note
that the lattice sizes are still too small to completely push the
oscillations to the origin. The agreement is, however, quite good.

For hard-core bosons (the $\delta$-function gas of Ref.\ \onlinecite{ll63}),
the one-particle density matrix $\rho(r)$ may be either
deduced in terms of generalizations of the Painlev\'{e} transcendent
\cite{vt79,jmms80}, by application of the quantum inverse scattering
method \cite{iikv92} using the equivalence with the non-linear
Schr\"{o}dinger equation, or by random matrix theory \cite{su92}.
In all methods the critical exponent is found to be $\eta=1/2$.
Data taken from Ref.\ \onlinecite{jmms80} again shows good agreement
with the above scaling law in Fig.~\ref{fig-hc}.

Finally, the long-ranged $g/r^{2}$ models have been shown to have a set of
critical exponents compatible with conformal predictions, both for their
discrete versions such as the Haldane-Shastry model \cite{gv87,s88,ha88}
and for the continuum system \cite{ky91,ko91,mz91,rs92}.
The critical exponents vary continuously with interaction strength $g$
and $\eta_{g}$ is independent of density $d$.
In Fig.~\ref{fig-rs} we plot the results of Ref.\ \onlinecite{rs92}
for $\rho(r)$ and the scaling curve corresponding to (\ref{eqn-etcf}).
Note that the $\eta_{2}=1$
curve corresponds to the Haldane-Shastry spin-chain. In addition, the
continuum data coincides with the discrete data, the first being the
low-density limit of the second.

Another important consequence following from (\ref{eqn-etcf}) is that
Monte-Carlo simulations to determine critical exponents can now
be restricted to a single calculation for each lattice size $L$.
This is done most conveniently by studying the behavior half way around
the chain. So, using as an example the spin-correlator in the $s=1/2$
antiferromagnet,
\begin{equation}
\FL
\langle S_{0} S_{L/2} \rangle \propto
 \left(
  \frac{ 2 \pi }{ L }
 \right)^{\eta}.
\label{eqn-hwetcf}
\end{equation}
Taking data from Kubo et al \cite{kkb88}, we show a plot
of $\langle S_{0} S_{L/2} \rangle$ versus $1/L$ in Fig.~\ref{fig-eta}.
A least-squares fit
gives $\eta=.94787$ which is a reasonable fit for just 5 data points,
especially in light of the logarithmic corrections.
Note that the errorbars are estimates of the oscillation amplitudes and
do not represent the statistical Monte-Carlo errors.
Incidentally, this method has been used by the present authors to compute
the critical exponents in the $g/r^{2}$ models to high accuracy \cite{rs92}.

Let us close this short note with a speculation. Using our interpretation
of equation (\ref{eqn-cf}), we may compute the time correlators at a single
site by choosing $u_{1}-u_{2} = t$, $v_{1}-v_{2}=0$.
Surpressing the $v_{1}, v_{2}$ arguments, the result is given by
\begin{displaymath}
\FL
\langle \phi(t),\phi(0) \rangle =
 \left(
  \frac{ 2 \pi }{ L }
 \right)^{2x}
 \left(
  \frac{ 1 }{ \sinh \frac{\pi t}{L} }
 \right)^{2x}.
\end{displaymath}
By choosing $v_{1}-v_{2}= L/2$, the same reasoning gives the large distance
asymptotics of the time correlator,
\begin{displaymath}
\FL
\langle \phi(t),\phi(0) \rangle =
 \left(
  \frac{ 2 \pi }{ L }
 \right)^{2x}
 \left(
  \frac{ 1 }{ \cosh \frac{\pi t}{L} }
 \right)^{2x}.
\end{displaymath}
In Fig.~\ref{fig-ltld} we plot the complete behavior of (\ref{eqn-cf})
for hard-core bosons in the $(t,r)$-plane.
It will be very interesting to compare these predictions to the calculations
for the hard-core bose gas of Ref.\ \onlinecite{iikv92} and for the
$\sigma$-model of Ref.\ \onlinecite{a93}.

%\acknowledgements

% figures

\figure{
\caption{
The scaled and normalized spin-correlator for $N=24,30,40$.
The line is the predicted scaling behavior for $\eta=1$.
Note that even the nearest-neighbor correlations are correctly predicted.}
\label{fig-af}}

\figure{
\caption{
The lattice version of the one-particle density matrix expansion $\rho(r)$ for
hard core bosons is compared to data from a direct calculation.
\label{fig-hc}}

\figure{
\caption{
The scaled and normalized one-particle density function for the $g/r^{2}$
model is plotted with the predicted scaling curves.
{}From left to right, the curves correspond to
 $\eta=5/4$ ($g=-1/2$ fermionic, $g=4$ fermionic),
 $\eta=1$ ($g=0$ fermionic, $g=4$ bosonic),
 $\eta=1/2$ ($g=0$ bosonic),
 $\eta=1/4$ ($g=-1/2$ bosonic)
and are symmetric about the y-axis.}
\label{fig-rs}}

\figure{
\caption{
We plot the spin-correlator halfway around the lattice, i.e.\
 $\langle S_{0} S_{N/2}\rangle$, as a function of the inverse lattice
length $1/N$.
Data for $N=6, 12, 24, 32, 40$ is taken from the Monte-Carlo calculation of
Kubo et al.
The solid curve is a fit for the Luther-Peschel $r^{-\eta}$
result which gives the estimate $\eta=.94787$.}
\label{fig-eta}}

\figure{
\caption{
The scaled correlation function of hard-core bosons ($\eta=1/2$)
in the $(t,z)$-plane. The value at $t=0$, $r=L/2$ is normalized to $1$.}
\label{fig-ltld}}

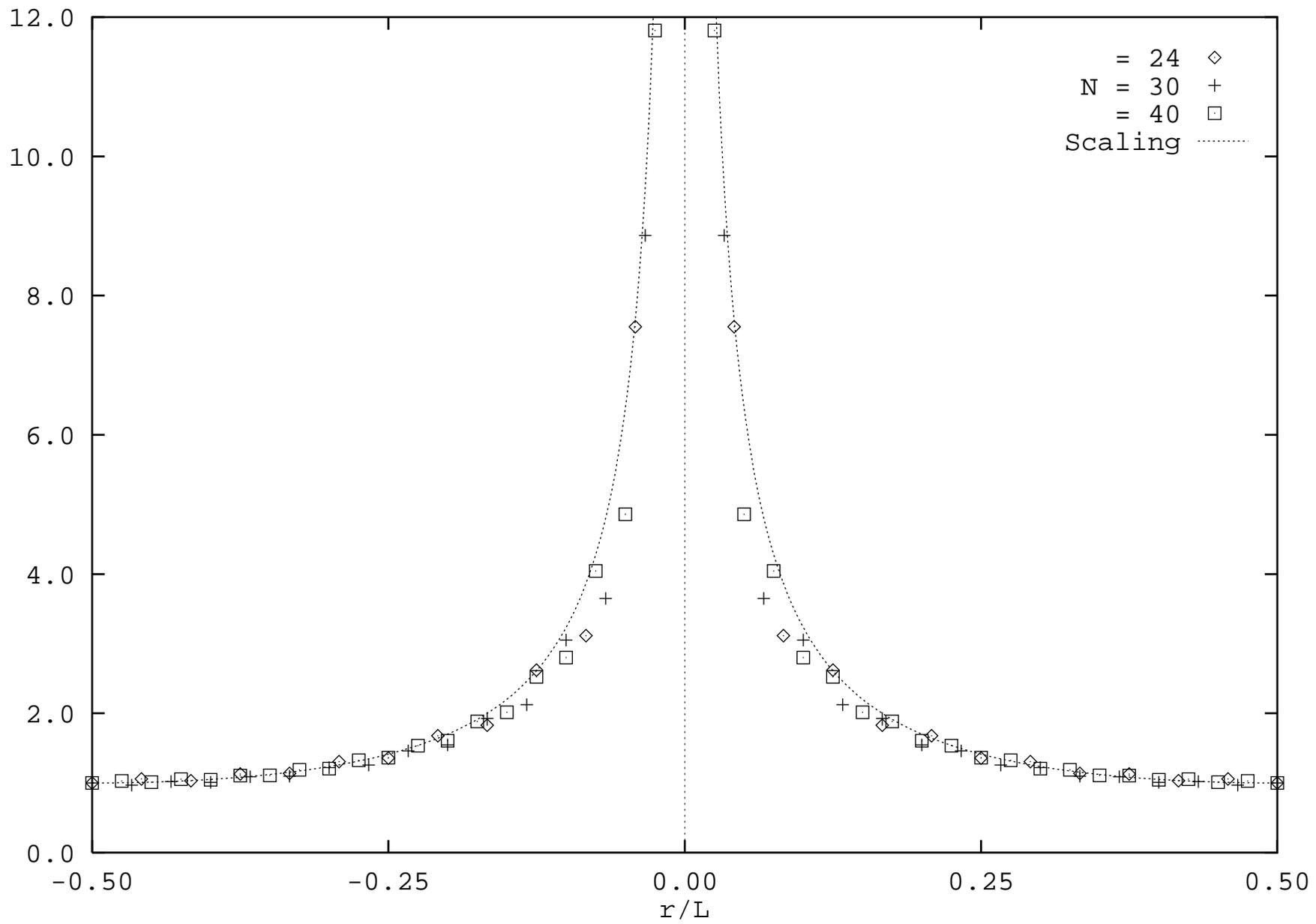
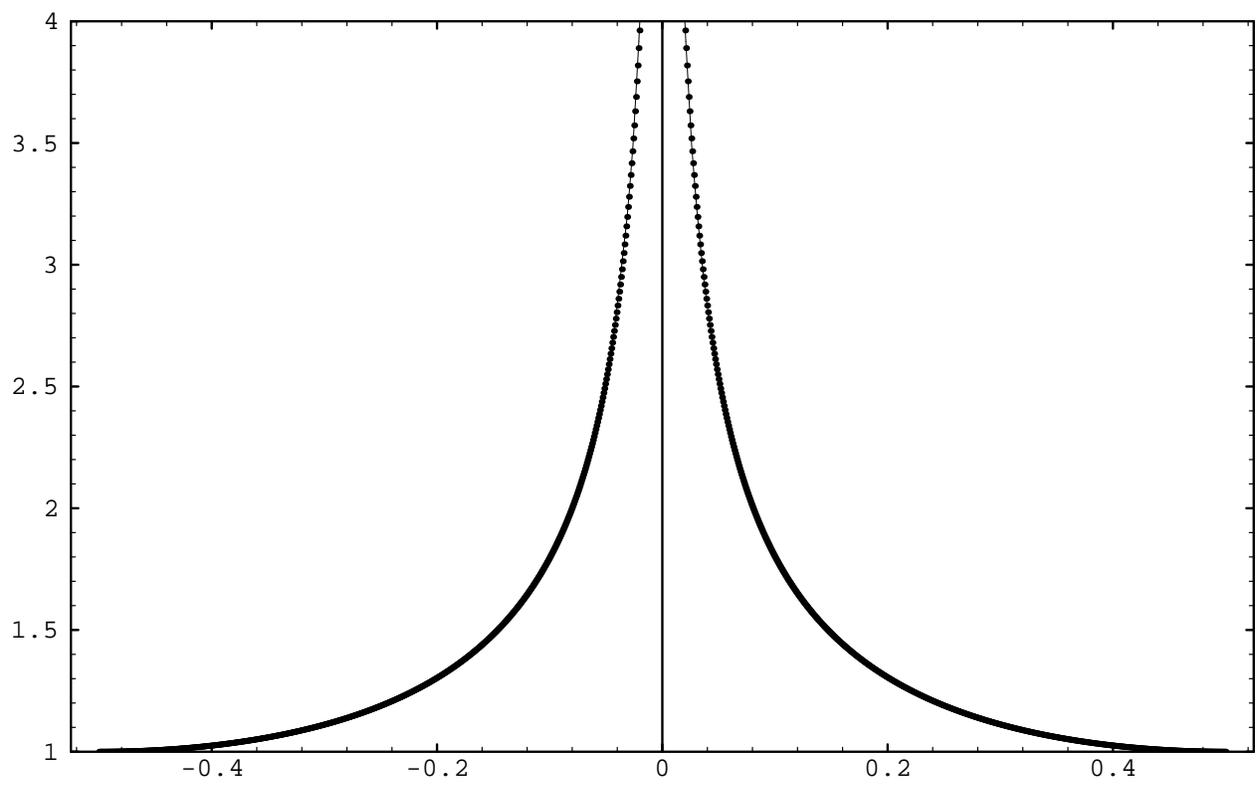
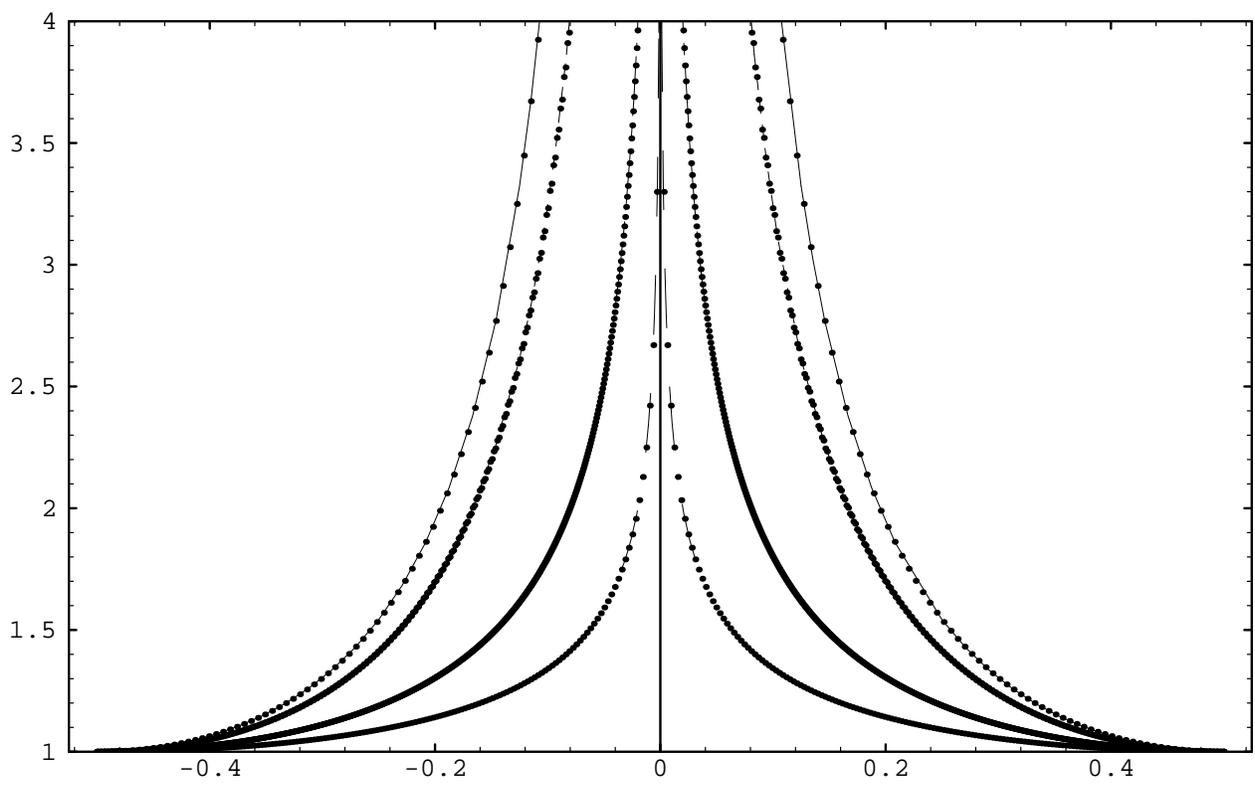
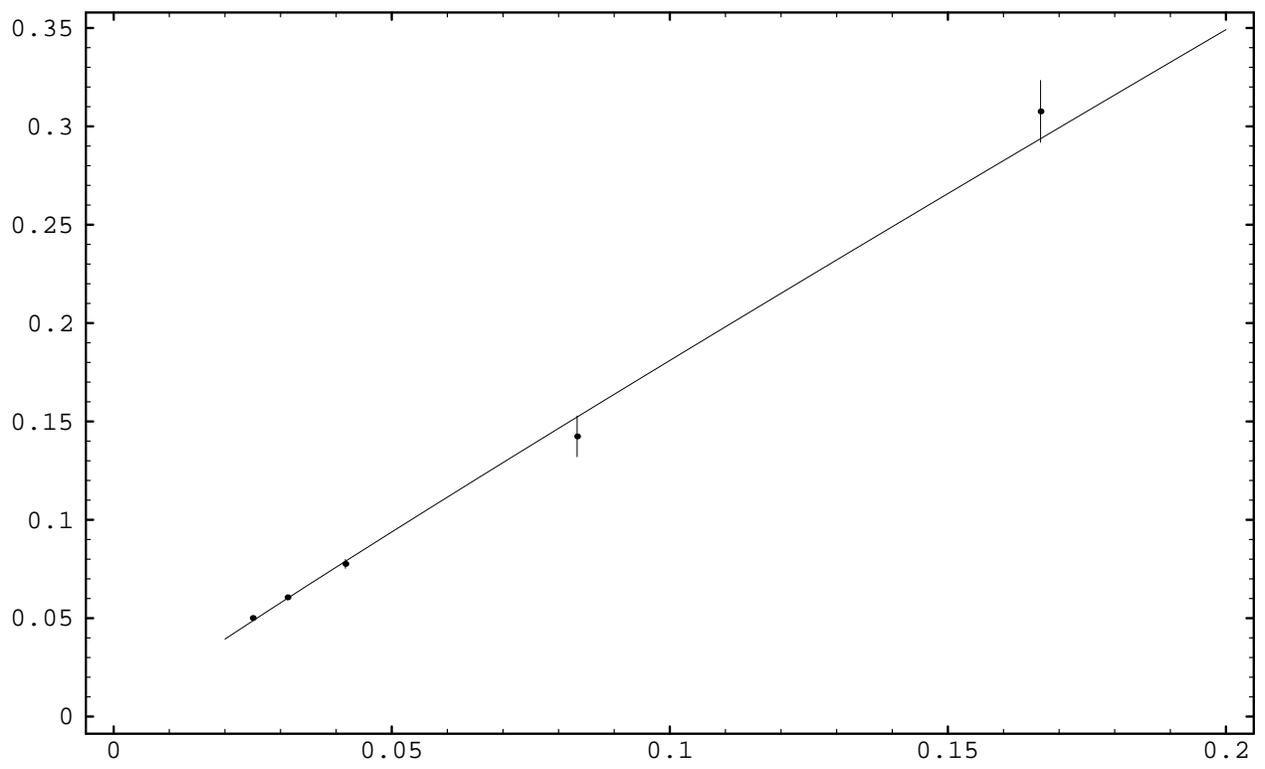
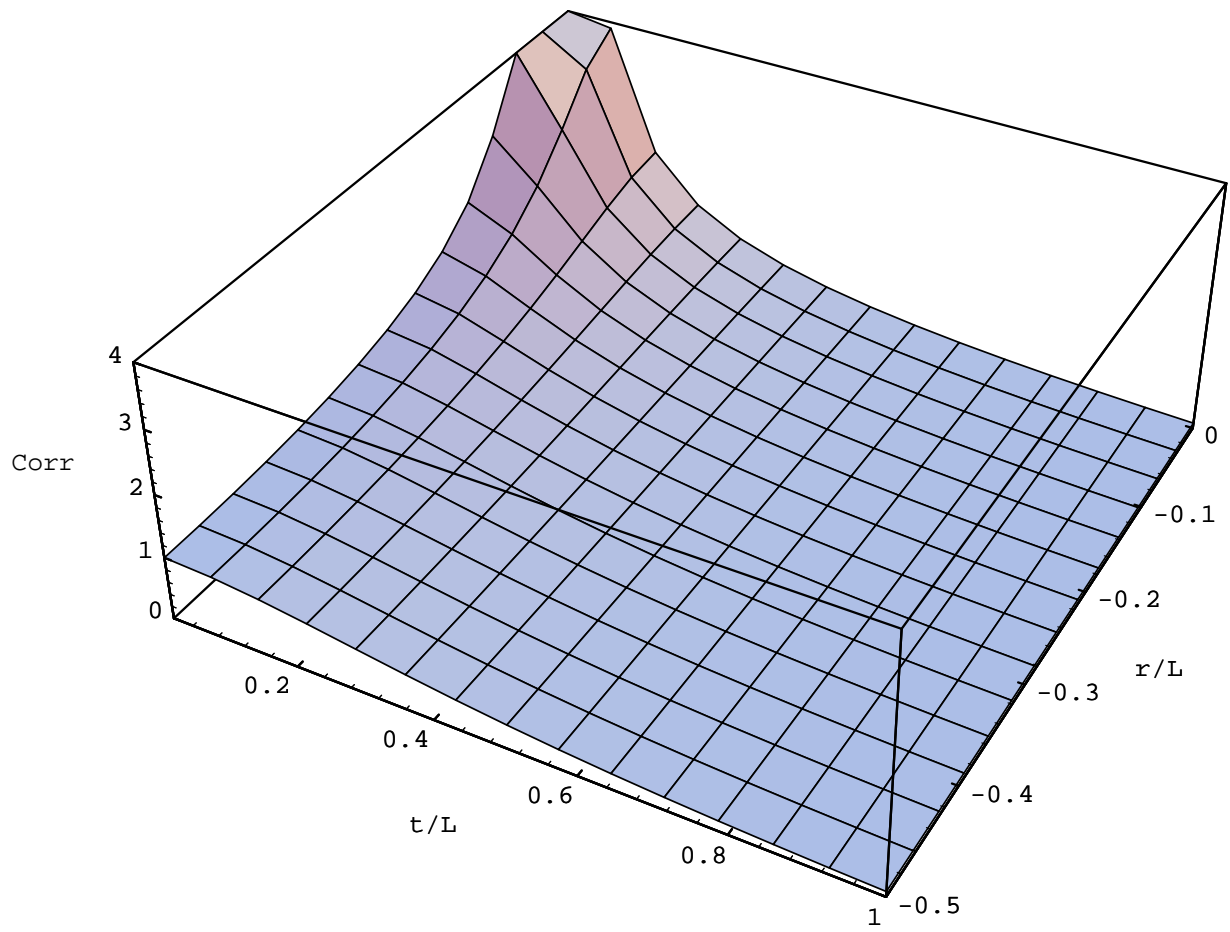
\end{document}